\documentclass{article}
\usepackage{graphicx}
\usepackage{amsmath}

\input{tcilatex}

\begin{document}

\title{Einstein's relativistic Doppler formula}
\author{Dr. Yves Pierseaux \\
Universit\'{e} Libre de Bruxelles, Physique des particules
(ypiersea@ulb.ac.be)}
\maketitle

\begin{abstract}
We note that Einstein's relativistic Doppler formula presents a strange
aspect.\ For incident light received under a fixed non zero angle, the
Doppler shift will change from blueshift to redshift (or vice-versa) for
some critical relative velocity of the observer and the source.
\end{abstract}

\begin{quote}
\bigskip It is very well known that Einstein's deduction of doppler
relativistic formula is entirely based on the invariance of the phase of a
planewave. Einstein's obtains the following formula (1):
\end{quote}

\begin{quotation}
\textit{From the equation for }$\omega ^{\prime }$\textit{\ it follows that
if an observer is moving with velocity v relatively to an infinitely distant
source of light of frequency }$\nu ,$\textit{\ in such a way that the
connection line ''source-observer'' makes an angle }$\phi $\textit{\ with
the velocity of the the observer referred to a system of co-ordinates which
is at rest relatively to the source of light, the frequency }$\nu ^{\prime }$%
\textit{\ of the light perceived by the observer is given by the equation}

\begin{equation}
\nu ^{\prime }=\nu \frac{1-\frac{v}{c}\cos \phi }{\sqrt{1-\frac{v^{2}}{c^{2}}%
}}
\end{equation}
\textit{This is Doppler principle for any velocities whatever. When }$\phi
=0,$\textit{\ the equation assumes the perspicuous form}

\textit{\bigskip }

\begin{equation}
\nu ^{\prime }=\nu \sqrt{\frac{1-\frac{v}{c}}{1+\frac{v}{c}}}
\end{equation}

\textit{We see that in contrast with the customary view, when }$%
v=-c,v^{\prime }=\infty $

\textit{Now if we put in Einstein's formula }$\phi =\frac{\pi }{2}$\textit{\
in (1), we obtain}

\textit{\bigskip } 
\begin{equation}
\nu ^{\prime }=\nu \frac{1}{\sqrt{1-\frac{v^{2}}{c^{2}}}}
\end{equation}
\end{quotation}

The immediate constatation is that for $\phi =0$ (2) we have a redishift ($%
\sqrt{\frac{1-\frac{v}{c}}{1+\frac{v}{c}}}<1)$ and for $\phi =\frac{\pi }{2}$
we have a blueshift ($\frac{1}{\sqrt{1-\frac{v^{2}}{c^{2}}}}>1)$

It is clear that for each velocity v, there is angle for which there is no
Doppler effect ($D=1$). It is very easy to find the critical angle $\phi _{t}
$:

With the Doppler factor
\begin{equation}
D=\frac{1-\frac{v}{c}\cos \phi }{\sqrt{1-\frac{v^{2}}{c^{2}}}}=1
\end{equation}

we obtain 
\begin{equation}
\cos \phi _{t}=\frac{1-\sqrt{1-\frac{v^{2}}{c^{2}}}}{\frac{v}{c}}\approx 
\frac{1}{2}\frac{v}{c}
\end{equation}

In standard books on special relativity, we always find the non-null
transverse effect but never this null effect for $\phi _{t}$ $\approx \frac{1%
}{2}\frac{v}{c}$. The value of $\cos \phi $ is decreasing from $\cos \phi =1$
($\frac{v}{c}\rightarrow 1$ $,\phi =0)$ to $\cos \phi =0$ $(\frac{v}{c}%
\rightarrow 0$ $,$ $\phi =90%
{{}^\circ}%
).$ We find for the small velocities that the critical\textit{\ angle of
transition }$\phi _{t}$ \textit{between blueshift and redshift} is a right
angle (change of sign of cosinus or change of sign of the velocity).

We obtain inversely from (4) 
\begin{equation*}
\frac{v}{c}=\frac{2\cos \phi _{t}}{1+\cos ^{2}\phi _{t}}
\end{equation*}

So for example if $\cos \phi _{t}=\frac{1}{2},$ $\frac{v}{c}=\frac{4}{5}.$
So for an angle of 60$%
{{}^\circ}%
$ and for a velocity about 240.$000$ \ km/s, Einstein's relativistic formula
predicts that there is \textbf{no Doppler effect}. If we take the angle $%
\phi ^{\prime }$ in the system of observer with the formula $\cos \phi
^{\prime }=\frac{\cos \phi -\frac{v}{c}}{1-\frac{v}{c}\cos \phi }$ , we
arrive exactly to the same conclusion. We note that for $\phi ^{\prime }=90%
{{}^\circ}%
,$ we obtain $\cos \phi =\frac{v}{c}$ $\neq \cos \phi _{t}\approx \frac{1}{2}%
\frac{v}{c}.$ 

\textit{So the critical angle }$\phi _{t}$\textit{\ has nothing to do with
the relativistic transformation of angle.}

There is no explanation for this very singular transition blueshift-redshift
in the framework of Einstein's kinematics. So without any change of the
orientation of the velocity (for example a receding object) we pass from a
redshift to a blueshift by a change of the intensity of the velocity. 

We will prove in another communication that this very serious problem is
connected with \textit{Poincar\'{e}'s elongated ellipse} from which we
obtain another definition of the units of space and time and also \textit{%
another relativistic Doppler formula} (reference 2).

References

\bigskip (1) Einstein A \ ''Zur Elektrodynamik bewegter K\"{o}rper'', Ann.
d. Ph., 17, p892-921, 1905, (we use English translation in ''The Principle
of relativity'', introduction and comment by Sommerfeld, Dover, New York,
p37-65, 1952).

(2) Pierseaux Y ''Einstein's spherical light waves versus Poincar\'{e}'s
ellipsoidal light waves'', London, September 2004, \`{a} para\^{i}tre dans
les annales de la fondation de Broglie.

\bigskip

\bigskip

\end{document}